# Deep Learning Accelerated First-Principles Quantum Transport Simulations at Nonequilibrium State

Zili Tang[1], Xiaoxin Xie[1], Guanwen Yao[1], Ligong Zhang[1], Xiaoyan Liu[1,2], Xing Zhang[1,2], and Liu Fei[1,2*]

[1]*School of Integrated Circuits, Peking University, Beijing, 100871, China.*

[2]*Beijing Advanced Innovation Center for Integrated Circuits, Beijing, 100871, China*

[*]To whom correspondence should be addressed.
E-mail: (F. L.) feiliu@pku.edu.cn.

**Abstract**：

The non-equilibrium Green's function method combined with density functional theory (NEGF-DFT) provides a rigorous framework for simulating nanoscale electronic transport, but its computational cost scales steeply with system size. Recent artificial intelligence (AI) approaches have sought to accelerate such simulations, yet most rely on conventional machine learning, lack atomic resolution, struggle to extrapolate to larger systems, and cannot predict multiple properties simultaneously. Here we introduce DeepQT, a deep-learning framework that integrates graph neural networks with transformer architectures to enable multi-property predictions of electronic structure and transport without manual feature engineering. By learning key intermediate quantities of NEGF-DFT—the equilibrium Hamiltonian and the non-equilibrium total potential difference—DeepQT reconstructs Hamiltonians under both equilibrium and bias conditions, yielding accurate transport predictions. Leveraging the principle of electronic nearsightedness, DeepQT generalizes from small training systems to much larger ones with high fidelity. Benchmarks on graphene, $MoS_2$, and silicon diodes with varied defects and dopants show that DeepQT achieves first-principles accuracy while reducing computational cost by orders of magnitude. This scalable, transferable framework advances AI-assisted quantum transport, offering a powerful tool for next-generation nanoelectronic device design.

# Introduction

The accurate simulation of quantum transport[1] in nanoscale devices is paramount for the advancement of next-generation electronics, optoelectronics, and energy conversion technologies. The Non-Equilibrium Green's Function (NEGF) formalism[2] coupled with Density Functional Theory (DFT)[3,4], commonly known as NEGF-DFT[5], stands as a powerful first-principles method, offering profound insights into the electronic structure and transport properties of materials and devices at the atomic scale. Its ability to self-consistently model open quantum systems far from equilibrium has made it an indispensable tool for designing and understanding phenomena in molecular junctions[6], nano-transistors[7], thermoelectric materials[8], and spintronic devices[9].

Despite its predictive power and detailed insights, the practical application of NEGF-DFT is often hampered by its substantial computational cost. The iterative solution of the Dyson and Keldysh equations[10,11], coupled with the self-consistent determination of the charge density, involves computationally intensive matrix inversions and diagonalizations. This computational burden scales unfavorably with system size, typically as $O(N^3)$. Consequently, large-scale simulations of realistic device architectures or high-throughput screening of novel materials often become prohibitively expensive, limiting the scope and pace of discovery.

To overcome these computational bottlenecks, machine learning (ML)[12] and particularly deep learning (DL)[13] techniques have emerged as a promising avenue. The ability of deep neural networks to learn complex, high-dimensional relationships directly from data offers the potential to significantly accelerate computationally demanding tasks in materials science[14] and quantum chemistry[15]. In the context of NEGF-DFT, DL models could learn to predict key quantities such as the Hamiltonian, Green's functions, or the self-consistent density matrix, thereby bypassing or accelerating the most expensive steps of the conventional calculation. However, effectively applying deep learning to the intricacies of NEGF-DFT calculations presents unique challenges, requiring neural network architectures specifically designed to capture the underlying physics and symmetries of quantum mechanical systems.

Here, we present DeepQT, a novel AI-accelerated framework for quantum transport prediction that combines graph neural networks[16,17] with Transformer[18,19] architectures, aiming to provide a significant leap in the speed and applicability of NEGF-DFT calculations. Rather than learning final physical quantities directly, DeepQT predicts key intermediate variables in the non-equilibrium Green's function combined with density functional theory (NEGF-DFT) formalism—specifically, the equilibrium Hamiltonian and the non-equilibrium total potential difference (TPD). Using SIESTA/TranSIESTA[20-23] as the reference first-principles platform, we decompose the full NEGF-DFT Hamiltonian into equilibrium (zero-bias) and non-equilibrium (biased) components. The model initially predicts the equilibrium Hamiltonian, followed by the TPD under an applied bias, from which the Hamiltonian correction is computed via integration over the basis functions. This correction is combined with the equilibrium Hamiltonian to recover the non-equilibrium Hamiltonian, which serves as input for quantum transport solvers to compute transmission spectra, density of states, and current–voltage characteristics.

To enhance generalizability, DeepQT leverages the electronic nearsightedness principle[24,25], allowing the model to be trained on small-scale systems and deployed for accurate predictions on significantly larger devices. We demonstrate its effectiveness across representative systems—including graphene, $MoS_2$, and silicon—featuring a range of defect and doping configurations. Our results show that DeepQT achieves first-principles accuracy in predicting both electronic structure and quantum transport properties, while significantly reducing computational cost. This approach provides a powerful and scalable solution for accelerating first-principles quantum transport simulations and offers a promising foundation for the design and analysis of nanoelectronic devices at advanced technology nodes.

## Results and Discussion

### Theoretical framework of DeepQT

In the NEGF-DFT framework, self-consistency of the Hamiltonian, electronic density, and potential is achieved by iteratively solving the Green's function and Poisson equations. The resulting Hamiltonian retains the Kohn–Sham form, but the electron density is obtained from the Green's function. However, this process becomes computationally prohibitive for large-scale devices, especially under non-equilibrium bias, where the cost can exceed that of equilibrium calculations by an order of magnitude. To address this inefficiency, we develop an AI-based framework that bypasses iterative self-consistency while enabling accurate multi-property predictions. The theoretical foundation of this model is presented below.

Under an applied bias, the Hamiltonian for an open device system takes the form[5,21]:

$$\left[-\frac{1}{2}\nabla^2 + V^{PS}(\mathbf{r}) + V_H(\mathbf{r}) + V_{xc}(\mathbf{r}) + V_{ext}(\mathbf{r})\right]\phi_i(\mathbf{r}) = \varepsilon_i\phi_i(\mathbf{r}), \quad (1)$$

Here, $V_{ext}(\mathbf{r})$ denotes the external potential induced by the applied bias, encompassing both the Hartree potential and exchange–correlation potential corrections arising from changes in electron density. Both the pseudopotential $V^{PS}(\mathbf{r})$ and $V_{ext}(\mathbf{r})$ are functions of the atomic positions, and the resulting electron density distribution is similarly dependent on the atomic coordinates. Consequently, the Hamiltonian under bias can be viewed as a mapping from the atomic coordinates $R$ and applied bias $V_b$ to the NEGF-DFT Hamiltonian:

$$\{R, V_b\} \rightarrow H_{NEGF-DFT}\{R, V_b\}, \quad (2)$$

This bias-dependent Hamiltonian is substantially more intricate than its equilibrium counterpart, as it is entangled with the external potential and necessitates a computationally intensive self-consistent solution.

Under non-equilibrium conditions with an applied bias voltage, both the electron density and potential distributions deviate from their equilibrium states, and the symmetry of the Hamiltonian is broken. Given that the applied bias is typically small (ranging from a few millivolts to a few volts), the biased Hamiltonian

$H_{NEGF-DFT}\{R, V_b\}$ can be approximated as the sum of the equilibrium Hamiltonian and a Hamiltonian correction term arising from the applied bias[22]:

$$H_{NEGF-DFT}\{R, V_b\} = H_{eq}\{R\} + \Delta H_{neq}\{R, V_b\}, \tag{3}$$

$$H_{eq}\{R\} = -\frac{1}{2}\nabla^2 + V^{PS}(r) + V_H(r) + V_{xc}(r), \tag{4}$$

$$\Delta H_{neq}\{R, V_b\} = V_{ext}(r), \tag{5}$$

Here, $H_{eq}$ denotes the equilibrium Hamiltonian without bias, which is solely a function of the atomic coordinates $R$. The term $\Delta H_{neq}$ represents the total potential difference (TPD) between the equilibrium and non-equilibrium states and depends on both the atomic coordinates $R$ and the applied bias $V_b$. Given $H_{eq}$, an additional prediction of $\Delta H_{neq}$ is required to bypass the time-consuming self-consistent calculation under bias. The Hamiltonian correction matrix can then be obtained by integrating over the basis functions (numerical atomic orbitals):

$$\Delta H_{V_b}^{ia,jb}\{R\} = \int C_{ia}^* \phi_{ia}^*(r) \Delta H_{neq}\{R, V_b\} C_{jb} \phi_{jb}(r) dr, \tag{6}$$

Here, $\phi_{ia}^*(r)$ denotes the conjugate of the $a$-th basis function of atom $i$, and $\phi_{jb}(r)$ denotes the $b$-th basis function of atom $j$. The terms $C_{ia}^*$ and $C_{jb}$ represent the corresponding weight coefficients of the basis functions.

To overcome this challenge, we introduce **DeepQT** (Figure 1a), an AI-based framework that replaces the computationally intensive self-consistent procedure. In this approach, the open device system is partitioned into the left and right electrodes (L and R), the central device region (D), and an applied bias $V_b$. The objective is to accurately predict the device Hamiltonian under various bias conditions, enabling efficient computation of multiple transport properties via quantum transport solvers **TBtrans**.

The DeepQT architecture comprises two sub-models, DeepQTH and DeepQTV, as shown in Figure 1b. DeepQTH predicts the equilibrium Hamiltonian matrix $H_{eq}^{ia,jb}\{R\}$ in the absence of bias, while DeepQTV predicts the Hamiltonian correction matrix $\Delta H_{V_b}^{ia,jb}\{R\}$ under various bias conditions. The combined output of these two models yields the full Hamiltonian matrix under bias:

$$H_{V_b}^{ia,jb}\{R\} = H_{eq}^{ia,jb}\{R\} + \Delta H_{V_b}^{ia,jb}\{R\}, \tag{7}$$

When calculating the Hamiltonian correction matrix, the choice of basis functions is critical for ensuring both computational efficiency and accuracy. We employed slater-type orbitals (STOs), whose numerical form remains fixed during integration and model training. Crucially, to generalize the DeepQT model to large-scale systems, we trained a mapping from local atomic structures to Hamiltonian blocks and TPD values by leveraging the electronic nearsightedness principle, thereby avoiding the influence of environments beyond the nearsightedness range.

To predict quantum transport properties, it is typically sufficient to compute the Hamiltonian at a limited number of bias points and then accelerate the evaluation of transport characteristics through interpolation. In the calculation workflow, the electrode Hamiltonian is first obtained, followed by the Hamiltonian of the entire open system. As electrodes are generally composed of conductive materials and are relatively small in size, their Hamiltonians can be computed directly using SIESTA with minimal computational cost. For semiconductor electrodes or large electrodes with lateral periodicity, Bloch's theorem and $k$-point sampling[22] can be employed to calculate only the smallest unit cell, thereby reducing both storage and computational demands. By combining the Hamiltonian of the open device system with that of the electrodes and inputting them into the TBtrans program, a wide range of transport properties can be efficiently computed.

When performing transport calculations on large-scale devices using the same electrode structure, k-point sampling, and energy grid as those of a smaller system, the precomputed surface Green's function files (*.TSGF**) from the smaller system can be directly reused, substantially reducing computational cost and improving efficiency. For systems with different electrode structures, if the *.TSGF** files are not provided and the '*out-of-core*' parameter is set to true (default), TBtrans automatically reconstructs the surface Green's functions from the electrode *.TSHS* files and computes the self-energies by incorporating the coupling terms between the scattering region and electrodes. Although this approach entails a higher computational cost, it ensures both the correctness and reproducibility of the self-energy calculation without requiring user intervention.

**Neural network architecture of DeepQTH**

In the absence of an applied bias, the system remains in equilibrium, and the converged Hamiltonian obtained from NEGF-DFT exhibits the same spatial distribution as that from standard DFT, differing only by minor numerical deviations. (see Supplementary Section 1). Accordingly, we draw on existing machine learning approaches developed for DFT Hamiltonian prediction[26-30] to predict the converged NEGF-DFT Hamiltonian under zero-bias equilibrium conditions.

To predict the equilibrium Hamiltonian of a large-scale system, we first estimate the blocks of the interatomic interaction Hamiltonian matrix $H_{eq}^{ia,jb}\{R\}$. Due to the rotation covariance in the global coordinate system, we transform $H_{eq}^{ia,jb}$ into the rotation-invariant form $H'_{iu,jv}$ defined in the local coordinate system (see Supplementary Section 2.1 for details on the construction of the local coordinate system in equilibrium). Leveraging the electronic nearsightedness principle, we focus solely on the atomic local environment, predict the Hamiltonian block $H'_{iu,jv}$, and then recover the final Hamiltonian in the global coordinate system through inverse rotation and block concatenation.

Inspired by the Graphformer[31,32] architecture we developed DeepQTH, a deep learning model with scalable node features is designed to predict the self-consistent converged Hamiltonian matrix at equilibrium. DeepQTH employs an attention mechanism to capture local atomic structural information and incorporates explicit structural and topological features; its global receptive field and adaptive aggregation strategy markedly enhance the model's expressive capacity.

We represent the local atomic structure as a graph model, with each atom $i$ as a node whose atomic number is $Z_i$. The bond length $|r_{i,k}|$ between atom $i$ and its neighbor atom $k$ is the edge feature $e_{i,k}$, $k \in N_i$, where $N_i$ represents the neighbor atom within the truncation radius of the atom $i$. The block of the Hamiltonian matrix $H'_{iu,jv}$ in the local coordinate system is represented by the edge feature of the output layer $e_{i,j}^{out}$. We define the atomic number feature embeddings $n_i^0$ and the Gaussian

feature embeddings $e_{i,k}^0$ of the bond lengths between nodes:

$$n_i^0 = \text{Embedding}(Z_i), \tag{8}$$

$$e_{i,k}^0 = \exp\left(-\frac{(|r_{i,k}|-r_n)^2}{2\sigma^2}\right), \tag{9}$$

$n_i^0 \in \mathbb{R}^d$, where $d$ denotes the user-defined dimensionality of the atomic node features. $e_{i,k}^0 \in \mathbb{R}^n$ represents the Gaussian expansion centered at each atomic coordinate $r_n$, where $\sigma$ is a parameter that controls the broadening of the Gaussian functions.

To improve the generalization capacity of the DeepQTH model in open systems containing defects, we incorporate local topological features of the nodes. In addition to atomic numbers, we introduce Laplacian centrality features $C_L(i)$, degree centrality features $C_D(i)$, and geometric features $\text{Vor}(i)$ derived from Voronoi diagrams (see Supplementary Section 3.1 for details on the three node-level topological features). These topological features help capture the influence of complex local environments near defects on the electronic structure, thereby enhancing the model's adaptability to defect-containing systems. Consequently, the initial node features $v_i^0$ are designed to be scalable, allowing topological attributes to be flexibly included or excluded depending on the specific task requirements:

$$v_i^0 = n_i^0 + C_L(i)W_1 + C_D(i)W_2 + \text{Vor}(i)W_3, \tag{10}$$

Here, $v_i^0 \in \mathbb{R}^d$, $W_1 \in \mathbb{R}^{1\times d}$, $W_2 \in \mathbb{R}^{1\times d}$ and $W_3 \in \mathbb{R}^{1\times d}$ are all learnable weight parameters.

Figure 2a illustrates the DeepQTH architecture, in which node embeddings and edge Gaussian embeddings are processed through the graph transformer layer (GTL) to predict the rotation-invariant Hamiltonian block $H'_{iu,jv}$ in the local coordinate frame (see Supplementary Section 3.3 for details of the GTL architecture). The output node features $v_i^{out}$, after layer normalization and dimensionality reduction, are used to visualize the atomic representations learned by DeepQTH.

In addition, the Hamiltonian can exhibit abrupt changes in response to minor structural perturbations, as demonstrated in previous studies[33,34]. Therefore, incorporating 3D spatial distance and spatial angular features is crucial for mitigating abrupt variations in the local coordinate system and effectively capturing the local atomic structural characteristics. To reduce the number of trainable parameters, DeepQTH introduces spatial angular features only at the output layer. The node and edge features in the output layer are updated as follows:

$$\widetilde{v_i^{l+1}}^{i,j} = GraphTransformerLayer\left(v_i^l, e_{i,k}^l, Y_{lm}(\theta_{i,k}^{i,j}, \varphi_{i,k}^{i,j})\right), \quad (11)$$

$$\widetilde{e_{i,j}^{l+1}}^{i,j} = Sigmoid\left(\left(\widetilde{v_i^{l+1}}^{i,j} \parallel \widetilde{v_j^{l+1}}^{i,j} \parallel e_{i,j}^l\right) W_4^{i,j} + b_4^{l+1}\right), \quad (12)$$

Where $\widetilde{v_i^{l+1}}^{i,j}$ and $\widetilde{e_{i,j}^{l+1}}^{i,j}$ represent the node $i$ features and the edge $r_{i,j}$ features at the $l$-th layer output defined in the local coordinate system on the edge $r_{i,j}$. Here, $\theta_{i,k}^{i,j}$ and $\varphi_{i,k}^{i,j}$ denote the polar and azimuthal angles of the vector $r_{i,k}$ defined in the local coordinate frame of the edge $r_{i,j}$. $Y_{lm}(\theta_{i,k}^{i,j}, \varphi_{i,k}^{i,j}) \in \mathbb{R}^N$ represent the spherical harmonics, where $N$ is the feature dimension, $l$ is the angular quantum number, and $m$ is the magnetic quantum number. $W_4^{i,j} \in \mathbb{R}^{(2d+n)\times no}$, $b_4^{l+1} \in \mathbb{R}^{no}$, $e_{i,j}^l \in \mathbb{R}^n$, $\widetilde{e_{i,j}^{l+1}}^{i,j} \in \mathbb{R}^{no}$, and $no$ denotes the product of the number of orbitals selected for atom pair $i,j$. The final output node and edge features are:

$$v_i^{out} = \widetilde{v_i^{l+1}}^{i,j}, \quad e_{i,j}^{out} = \widetilde{e_{i,j}^{l+1}}^{i,j}, \quad (13)$$

$$v_i^{final} = LayerNorm(v_i^{out}), \quad (14)$$

$$H'_{iu,jv} = Reshape(e_{i,j}^{out}), \quad (15)$$

Here, $v_i^{final}$ represents the feature of atom $i$ extracted by DeepQTH, which can be used for downstream tasks such as defect or doping detection, classification, or clustering. $e_{i,j}^{out}$ denotes the output feature of edge $r_{i,j}$, which is reshaped to obtain the predicted Hamiltonian block $H'_{iu,jv}$.

We define the error loss function:

$$LOSS = MSE(H'_{iu,jv}, \hat{H}'_{iu,jv}), \qquad (16)$$

Here, $\hat{H}'_{iu,jv}$ is the Hamiltonian block label used for training. After predicting $H'_{iu,jv}$, it is then transformed back into the global coordinate system to obtain the Hamiltonian block $H_{eq}^{ia,jb}$, thereby constructing the complete zero-bias Hamiltonian.

**Neural network architecture of DeepQTV**

To accelerate self-consistent calculations under non-equilibrium conditions, we employ **DeepQTH** to predict the equilibrium Hamiltonian and **DeepQTV** to estimate the Hamiltonian correction induced by bias. When a bias is applied, the potential varies along the transport direction, breaking the symmetry of the correction term. To address this asymmetry, we utilize the electronic nearsightedness principle by constructing local coordinate system centered at each grid point (see Supplementary Section 2.2 for details on coordinate construction under bias). By transforming the global coordinates of each local substructure into its local frame, we extract, for every grid point $i$, the angular information of atoms within the cutoff radius—namely the polar angles $\theta_{i,k}^i$ and azimuthal angles $\varphi_{i,k}^i$—along with the atomic numbers $Z_i$ and the initial TPD values $V_i$.

The DeepQTV model can also be implemented using the GTL framework. We treat the mesh points as virtual nodes $v_i^{gird}$, which are connected to all neighboring atoms within the local substructure. The bond lengths $|r_{i,k}|$ between nodes are encoded as edge features $e_{i,k}$ using gaussian functions, and the TPD values $\Delta H_{V_b}^i$ at the grid points are produced as the model outputs. The input features comprise atomic number embeddings and initial TPD embeddings for each node (with $Z_i = 0$ for grid nodes), Gaussian basis edge features, and angular features of neighboring atoms in the local coordinate system, defined as:

$$v_i^0 = Embedding(Z_i) + Embedding(V_i), \tag{17}$$

$$e_{i,k}^0 = \exp\left(-\frac{(|r_{i,k}|-r_n)^2}{2\sigma^2}\right), \tag{18}$$

Here, $v_i^0 \in \mathbb{R}^d$ denotes the combined embedding of the atomic number and initial TPD, while $e_{i,k}^0 \in \mathbb{R}^n$ represents the Gaussian basis feature. In systems with defects, the topological features of atomic nodes can be further incorporated, whereas grid nodes do not require such extensions.

The network architecture of DeepQTV is shown in Figure 2b. Considering the symmetry breaking introduced by the applied bias, the angular information of nodes in the local coordinate system is fed into the first GTL layer, enabling each subsequent layer to incorporate both the 3D spatial distance features and angular features of node pairs. The feature learning process at the first layer(0) is as follows:

$$\overrightarrow{v_i^1}^i = GraphTransformerLayer\left(v_i^0, e_{i,k}^0, Y_{lm}(\theta_{i,k}^i, \varphi_{i,k}^i)\right), \tag{19}$$

$$\overrightarrow{e_{i,k}^1}^i = Sigmoid\left(\left(\overrightarrow{v_i^1}^i \parallel \overrightarrow{v_k^1}^i \parallel e_{i,k}^0\right)W_5^1 + b_5^1\right), \tag{20}$$

Here, $\overrightarrow{v_i^1}^i \in \mathbb{R}^d$ denotes the output features of node $i$ at the first layer in the local coordinate system defined at the grid node $i$, and $\overrightarrow{e_{i,k}^1}^i \in \mathbb{R}^n$ denotes the output features of edge $r_{i,k}$. Here, $W_5^1 \in \mathbb{R}^{(2d+n)\times n}$ and $b_5^1 \in \mathbb{R}^n$. The subsequent layer($l$) are updated as follows:

$$\overrightarrow{v_i^{l+1}}^i = GraphTransformerLayer\left(\overrightarrow{v_i^l}^i, \overrightarrow{e_{i,k}^l}^i\right), \tag{21}$$

$$\overrightarrow{e_{i,k}^{l+1}}^i = Sigmoid\left(\left(\overrightarrow{v_i^{l+1}}^i \parallel \overrightarrow{v_k^{l+1}}^i \parallel e_{i,k}^l\right)W_5^{l+1} + b_5^{l+1}\right), \tag{22}$$

Here, $W_5^{l+1} \in \mathbb{R}^{(2d+n)\times n}$, $b_5^{l+1} \in \mathbb{R}^n$, and $\overrightarrow{e_{i,k}^{l+1}}^i \in \mathbb{R}^n$.

After passing through the model output layer, we extract the features $v_i^{gird}$ of the mesh nodes $i$:

$$v_i^{gird} = \overbrace{v_i^{l+1}}^{i}, \tag{23}$$

The virtual nodes features are subsequently mapped to the TPD values at each grid point through fully connected layers, as detailed below:

$$\Delta H_{V_b}^i = SiLU(v_i^{gird}W_6^{out} + b_6^{out})W_7^{out} + b_7^{out}, \tag{24}$$

Here, $SiLU$ refers to the nonlinear activation function, $v_i^{gird} \in \mathbb{R}^d$, $W_6^{out} \in \mathbb{R}^{d \times d}$, $W_7^{out} \in \mathbb{R}^{d \times 1}$, $b_6^{out} \in \mathbb{R}^d$, $b_7^{out} \in \mathbb{R}^1$ are all learnable weight parameters. After predicting the TPD values $\Delta H_{V_b}^i$ for all grid points, the TPD distribution can be obtained as $\Delta H_{neq}\{R, V_b\} = \{\Delta H_{V_b}^i, i \in N_g\}$.

By integrating the TPD distribution with basis functions, the Hamiltonian correction term $\Delta H_{V_b}^{ia,jb}$ and the Overlap matrix $S^{ia,jb}$ can be obtained:

$$\Delta H_{V_b}^{ia,jb}\{R\} = \int C_{ia}^* \phi_{ia}^*(r) \Delta H_{neq}\{R, V_b\} C_{jb} \phi_{jb}(r) dr$$

$$= C_{ia}^* C_{jb} \int \phi_{ia}^*(r) \Delta H_{neq}\{R, V_b\} \phi_{jb}(r) dr, \tag{25}$$

$$S^{ia,jb}\{R\} = \int C_{ia}^* \phi_{ia}^*(r) C_{jb} \phi_{jb}(r) dr = C_{ia}^* C_{jb} \int \phi_{ia}^*(r) \phi_{jb}(r) dr, \tag{26}$$

Here, $C_{ia}^* C_{jb}$ can be obtained through basis function correction using the Overlap matrix calculated by TranSIESTA and equation (26).

We define the loss function as follows:

$$LOSS = MSE(\Delta H_{V_b}^i, \Delta \widehat{H}_{V_b}^i), \tag{27}$$

The loss function is designed to constrain the model to accurately predict the TPD values of the grid points under bias, with $C_{ia}^* C_{jb}$ serving as a correction factor to ensure that the basis function can integrate to generate an accurate overlap matrix, thereby ensuring the prediction accuracy of the Hamiltonian correction term.

Transport properties were calculated using a Hamiltonian interpolation approach, and the DeepQTV model was trained separately at only a few bias voltages to predict both the TPD and the Hamiltonian correction terms. Finally, the equilibrium Hamiltonian and the correction terms were combined to obtain the final Hamiltonian under different bias conditions.

**Electronic Structure Prediction**

The DeepQTH sub-models are designed to probe the electronic structure of materials under equilibrium conditions without applied bias. In this study, graphene, $MoS_2$, and Si were chosen as representative systems, and independent DeepQTH models were trained for each. To evaluate model generalization and practical applicability, two types of DFT datasets were constructed: defect-free small-scale unit cell datasets and small-scale unit cell datasets containing defects or dopants (as described in the Method-Datasets section).

The DeepQTH models were first trained on defect-free unit cells and then tested on large-scale pristine supercells. Supplementary Fig. S5 shows predictions for three disorder-free systems: a 15×15×1 graphene sheet (450 atoms), a 10×10×1 $MoS_2$ monolayer (300 atoms), and a 5×5×5 Si supercell (250 atoms). The predicted band structures and densities of states exhibit excellent agreement with reference SIESTA calculations, demonstrating the capability of DeepQTH to accurately capture the electronic structure of both two- and three-dimensional multi-component systems. Importantly, the approach circumvents computationally expensive self-consistent calculations for large-scale materials while accurately capturing their electronic structure.

Building on these results, we further evaluated the performance of DeepQTH on defective and doped datasets. To enhance the model's generalization across both pristine and defective systems while mitigating catastrophic forgetting, we employed a mixed-data training strategy, fine-tuning the pre-trained defect-free model (see Supplementary Material 4.2). This approach enabled the model to integrate new knowledge while retaining prior learning, achieving a stable form of continuous learning.

Figure 3 presents predictions of the electronic structure in large-scale supercells with complex defects and doping. Figure 3a shows a 12×12×1 graphene supercell (284 atoms) containing Stone–Wales defect SW(55-77), double vacancy defects $V_2$(555-777) and $V_2$(5555-6-7777)[35]; the predicted band structure and density of states closely match SIESTA calculations, even when $V_2$(555-777) and $V_2$(5555-6-7777) defects are in proximity. Figure 3b illustrates predictions for a 10×10×1 MoS$_2$ supercell (296 atoms) with four common defects[36]: single sulfur vacancy V$_S$, double sulfur vacancy V$_{S2}$, one molybdenum atom replacing one or two sulfur atoms Mo$_S$ and Mo$_{S2}$, where the model successfully captured the defect-induced localized states within the band gap. Figure 3c shows a 6×6×3 silicon supercell (215 atoms) with substitutional boron dopants B$_{Si1}$, substitutional phosphorus dopants P$_{Si2}$ and Schottky defect V$_{Si2}$[37], again demonstrating strong agreement with SIESTA.

As shown in Table S2 of the Supplementary, for the three defect/doping systems, the mean absolute error (MAE) of the predicted Hamiltonian was 8.56, 11.71, and 10.12 meV, respectively. Compared with the model without mixed-data fine-tuning, these values represent reductions of nearly six-, nine-, and twelve-fold, demonstrating a substantial improvement in predictive accuracy. Notably, the fine-tuned model retained strong performance on defect-free datasets, with MAE increasing by only ~1.72, 3.44, and 3.97 meV, respectively. Given that the magnitudes of the Hamiltonian elements lie within the electron-volt range, such increases are negligible and well within acceptable limits. Finally, we used the sisl package[38] to store the predicted Hamiltonians and calculate the corresponding Fermi levels (see Supplementary Table S3), confirming that the predicted Hamiltonians reliably capture the electronic filling levels.

To further assess the scalability of DeepQTH, we evaluated the average Hamiltonian MAE across three defective materials at varying system sizes (Supplementary Figure S6a) and compared the performance with SIESTA's computation times (Supplementary Figure S6b). The results show that although the MAE of DeepQTH increases slightly with system size, it remains well within acceptable limits. Importantly, DeepQTH offers a substantial computational advantage over SIESTA, with significantly reduced calculation times. As shown in Supplementary Figure S6(b), SIESTA exhibits approximately cubic scaling with system size, whereas DeepQTH demonstrates near-linear scaling. These findings highlight the efficiency of DeepQTH, with even greater computational advantages anticipated as system sizes increase or as higher-order basis sets and more stringent convergence criteria are adopted.

DeepQTH achieves minimal prediction error by integrating angular features and node topological characteristics into the GTL architecture. Supplementary Table S1 presents an ablation study evaluating the contributions of node, distance, and angular features when trained on the mixed dataset. The results demonstrate that incorporating spherical harmonic angular features markedly reduces prediction error, while the addition of three node topological features further enhances accuracy, particularly in the presence of defect configurations.

We further benchmarked DeepQTH against several state-of-the-art models for Hamiltonian prediction, including SchNorb[27], PhiSNet[28], DeepH[29], and PaiNN[39]. As shown in Supplementary Table S4, all models were trained on a mixed dataset and evaluated on large-scale material systems. DeepQTH consistently outperformed the other models, particularly on test structures involving defects or doping, achieving a marked reduction in MAE. These results highlight the advantage of integrating angular features and topological node descriptors into the model architecture. DeepQTH demonstrates strong generalizability and high accuracy in predicting electronic structures across diverse materials and defect/doping configurations. Its capability in accurately capturing equilibrium-state Hamiltonians provides a solid foundation for extending to non-equilibrium Hamiltonians and quantum transport predictions.

**Transport Property Prediction**

Having established that DeepQTH accurately predicts the equilibrium DFT Hamiltonian, the model was further fine-tuned using NEGF-DFT Hamiltonians from small-scale devices computed with TranSIESTA, enabling reliable inference of equilibrium NEGF-DFT Hamiltonians for larger systems. To evaluate the performance of DeepQTV, the model was benchmarked on predicting the TPD in defect-free large-scale armchair $MoS_2$(A-$MoS_2$) device under various bias voltages (see Supplementary Section 7). As shown in Supplementary Table S5, DeepQTV accurately reproduces the real-space TPD distribution across different biases, achieving a MAE below 0.01 eV at all grid points and relative MAE below 2%.

Subsequently, the overlap matrix was computed by integrating basis functions

(with their radial components illustrated in Supplementary Figure S7), and further refined using overlap integrals obtained from TranSIESTA. The TPD was then integrated with the basis functions and corrected by the product $C_{ia}^* C_{jb}$ to yield the Hamiltonian correction term. Supplementary Figure S8 presents the resulting correction terms under various biases and compares the predicted non-equilibrium Hamiltonians to the real values self-consistently calculated by using TranSIESTA. The results show excellent agreement, with coefficients of determination approaching the ideal value.

Furthermore, as shown in Supplementary Fig. S9, we compared the core-hour requirements of DeepQTV predictions with those of TranSIESTA calculations. While the TranSIESTA computational cost increases markedly with applied bias, DeepQTV maintains an essentially constant prediction time across all bias voltages, delivering nearly a 4-fold reduction in computational effort under finite bias.

Together, these results demonstrate that DeepQTV can robustly predict TPD distributions and corresponding Hamiltonian correction items for large-scale devices in non-equilibrium states. When integrated within the **DeepQT** framework, the combination of DeepQTH and DeepQTV enables efficient and accurate prediction of non-equilibrium Hamiltonians across a wide range of bias conditions.

**Prediction of Transport Properties of Large-Scale Nanoporous Graphene Nanoribbons**

To evaluate the predictive capabilities of **DeepQT** in complex, previously unseen scenarios, we applied the model to nanoporous graphene diode (NGD)[40,41] structure with large-scale defects absent from the training data. As illustrated in Fig. 4a, the device consists of left and right electrodes constructed from 10 × 1 orthogonal graphene supercell slabs, with nine slabs located in the scattering region. The central region contains a circular hole defect of diameter D = 8 Å, and the total structure comprises 422 carbon atoms.

We first predicted the equilibrium Hamiltonian of the NGD at zero bias using **DeepQTH**, and computed the corresponding transmission spectrum via **TBtrans** (Fig. 4b). While the ideal bulk graphene exhibits a characteristic stepwise transmission profile (The gray solid line in Fig. 4b), the introduction of the hole defect significantly reduces and modulates the transmission, consistent with the emergence of localized states that disrupt coherent transport. The predicted transmission spectrum closely matches reference NEGF-DFT calculations, maintaining low error across the full energy window from −1.5 eV to +1.5 eV. This agreement underscores the model's capacity to capture both the band structure and defect-induced scattering effects. Notably, the model accurately generalizes to macroscopic defects never encountered during training, highlighting its robust extrapolation capability.

Fig. 4c compares the predicted and computed spectral density(ADOS) of the left electrode and Green's function density of states (DOS), revealing near-perfect correspondence, with only minor deviations at a few resonant peaks. This confirms DeepQT's fidelity in reproducing complex interfacial electronic structure and transport-relevant spectral features. Fig. 4d presents the atom-resolved DOS at the Fermi level ($E_F$=0.0 eV), showing strong localization of electronic states at the defect edges—primarily on undercoordinated atoms—and gradual restoration of uniform DOS further from the pore. This spatial distribution is characteristic of graphene systems and is essential for understanding band modulation through defect engineering.

In Fig. 4e, we compare the DOS among edge atoms, first-nearest neighbors (N-edge), and second-nearest neighbors (NN-edge). The DOS exhibits a clear spatial decay with distance from the defect center, consistent with Fig. 4d, further validating the model's capability to resolve local electronic features. Finally, Fig. 4f shows that the predicted current–voltage (*I–V*) characteristics under applied bias closely match reference calculations, confirming the model's accuracy in capturing non-equilibrium quantum transport.

We further applied DeepQT to predict transport properties of NGD with varying pore diameters (see Supplementary Section 9). The results reveal a non-monotonic dependence of transport performance on pore size—strongly influenced by bias voltage, available conduction channels, and the energy-dependent structure of the transmission spectrum. This complex behavior, accurately captured by DeepQT, underscores the model's utility for rapid and reliable analysis of defect-engineered quantum transport in graphene-based devices, paving the way for data-driven optimization of next-generation nanoscale electronic systems.

**Prediction of Transport Properties of silicon Esaki diodes**

To further evaluate the capability of DeepQT to predict quantum transport in doped large-scale devices, we selected a prototypical system with distinct transport characteristics—the silicon Esaki (tunnel) diode—as a stringent test case. The Esaki diode first reported by Leo Esaki[42] in 1957, the Esaki diode is formed by creating a heavily doped PN junction via rapid alloying such that the Fermi level lies within the conduction band of the n-type region and the valence band of the p-type region, rendering both sides degenerate[43]. The junction must also be sufficiently thin to allow electrons to tunnel directly from the n-type layer through the potential barrier into the p-type layer. Under forward bias, this device exhibits negative differential resistance at room temperature, with current dominated by quantum tunneling. Owing to these unique properties, the Esaki diode has been widely employed in low-noise high-frequency amplifiers, oscillators, and high-speed switching circuits[44-46].

We simulated a heavily doped silicon PN-junction Esaki diode, as shown in Fig. 5a, using a 2×2×12 cubic supercell comprising 384 atoms. The p-type region contains three boron dopants (green), and the n-type region three phosphorus dopants (orange), yielding an abrupt junction with a doping concentration of approximately $8 \times 10^{20}$ cm$^{-3}$.

The DeepQTH model was trained and subsequently fine-tuned on a combined dataset of pristine and defect-containing silicon structures (Supplementary Tables S8 and S11), whereas DeepQTV was trained on a smaller doped-silicon dataset (Supplementary Table S11) to predict TPD across a range of bias voltages and, in turn, to generate the corresponding equilibrium and nonequilibrium Hamiltonians. Quantum transport properties were then computed using **TBtrans**.

Fig. 5b presents the band structures of the left (p-type) and right (n-type) electrodes after Fermi-level alignment using SIESTA. In the p-type region, the bands shift upward, placing the valence-band maximum about 0.25 eV above the Fermi level and narrowing the gap to roughly 0.75 eV. In contrast, the n-type region shows a downward shift, with the conduction-band minimum about 0.25 eV below the Fermi level and a band gap of approximately 0.7 eV. The zero-bias local density of states (LDOS) (Fig. 5c) reveals a pronounced band offset and built-in potential barrier at the junction, with the Fermi level residing in the p-type valence band and the n-type conduction band—a hallmark of an Esaki diode. The junction width is about 10.86 Å, and the conduction-band minimum of the left electrode lies roughly 1.0 eV above the Fermi level of the right electrode.

After predicting the Hamiltonians at bias voltages from -1.0 to 1.0 V in 0.2 V increments using DeepQT, we input them into TBtrans to compute the *I-V* characteristics (blue points in Fig. 5d). The results closely match those from full TranSIESTA + TBtrans calculations (silver curve), capturing the nonlinear transport behavior. Notably, DeepQT reproduces the negative differential resistance (NDR) observed between 0.2 and 0.5 V forward bias, where the current decreases with increasing voltage—a key effect arising from band-structure evolution.

To probe the origin of the NDR behavior, Fig. 5e presents schematic diagrams of the LDOS under five different bias voltages (as indicated by the red dots in Figure 5d). At 0.0 V, the Fermi levels of the n- and p-type electrodes are aligned; the narrow junction allows limited electron tunneling between the n-type conduction band and the p-type valence band, but the net current is zero. At 0.25 V, upward band shifting in the n region enables electrons to tunnel unidirectionally into unoccupied states of the p-type valence band, opening a transport channel and generating a tunneling current. When the bias reaches 0.5 V, the n-type bands rise further, moving electrons out of the p-region's available states and into the band gap, thereby reducing the current and producing negative differential resistance (NDR). At 0.75 V, n-type carriers lie entirely within the p-type band gap, while the n-type Fermi level remains about 0.25 eV below the p-type conduction-band minimum and the p-type Fermi level about 0.2 eV above the n-type valence-band maximum, allowing electrons and holes a finite probability of crossing the barrier and generating a small current. At 1.0 V, the n-type Fermi level aligns with the p-type conduction-band minimum, establishing an efficient transport path and markedly increasing the current.

These results demonstrate that DeepQT accurately captures not only conventional transport behavior but also intricate nonlinear effects such as NDR—driven by complex band alignment and localized state interactions. This ability underscores its applicability to atomic-scale device design, particularly in systems where interfacial barriers and carrier control are critical. DeepQT provides an efficient and accurate framework for predicting electronic structure and quantum transport across equilibrium and non-equilibrium conditions. Its demonstrated robustness in handling doped and defective configurations establishes a new paradigm for scalable, data-driven device simulation, bridging the gap between first-principles fidelity and computational efficiency.

**Discussion**

In this work, we introduce **DeepQT**, a data-driven framework that accelerates quantum transport simulations by leveraging graph neural networks and Transformer architectures. DeepQT uniquely predicts both the equilibrium Hamiltonian and the non-equilibrium TPD, enabling efficient inference of bias-dependent quantum transport properties. The model incorporates a rich representation of atomic environments—combining node and edge features, 3D spatial distances, angular information and local topological features—and computes Hamiltonian correction items through basis function integrals. By integrating these components, DeepQT reconstructs the self-consistent non-equilibrium Hamiltonian without the need for iterative convergence, significantly reducing the computational cost of traditional NEGF-DFT simulations. Transport properties are subsequently computed using the **TBtrans** solver, bridging high-fidelity electronic structure predictions with efficient quantum transport analysis.

The innovation of DeepQT lies in its ability to unify equilibrium and non-equilibrium modeling within a single, scalable framework, enabling accurate simulations across diverse material systems and device configurations. Beyond analyzing electronic structures, DeepQT extends to predicting transport characteristics in large-scale, open-boundary devices—including those with complex defect or doping profiles—thus supporting integrated materials–device co-design at advanced technology nodes. As semiconductor devices scale further into the nanoscale regime, DeepQT offers a promising path forward for first-principles-informed device optimization. By circumventing the computational bottlenecks of self-consistent NEGF-DFT, it enables rapid exploration of materials and structures with physical rigor and computational efficiency.

Nevertheless, challenges remain. Accurately modeling systems with defects or dopants requires denser and more diverse training data to capture local perturbations. The inclusion of richer node features increases model complexity and training overhead. Moreover, the choice of truncation radius becomes critical in systems with composite defects or closely spaced dopants, where interference effects may degrade prediction accuracy. Despite these limitations, DeepQT marks a significant step toward scalable, AI-accelerated quantum transport simulation. Future efforts will aim to improve model generality and extend the framework to additional quantum observables, further advancing data-driven quantum device modeling.

**Method**

**Datasets**

To construct a comprehensive training and validation framework for DeepQT, we employed SIESTA/TranSIESTA to generate three categories of datasets: (1) defect-free periodic small-scale supercells, (2) periodic small-scale supercells with defects or dopants, and (3) small-scale open devices with localized defects or dopants. Each dataset spans three representative materials—graphene, monolayer $MoS_2$, and silicon—and is designed to systematically evaluate DeepQT's ability to predict both equilibrium electronic structure and non-equilibrium quantum transport properties.

**Defect-free periodic supercell datasets.** To assess model performance across different dimensionalities and elemental compositions, we generated random configurations of pristine structures via *ab-initio* molecular dynamics (MD) simulations at 300 K with a time step of 1 fs. Specifically, we considered $6 \times 6 \times 1$ monolayer graphene supercells, $5 \times 5 \times 1$ monolayer $MoS_2$ supercells, and $3 \times 3 \times 3$ silicon cubic supercells. From the MD trajectories, 600 energetically stable configurations were selected for each material and partitioned into training, validation, and test sets using a 6:2:2 split. All DFT calculations were performed using norm-conserving pseudopotentials[47] and the GGA-PBE functional[48], with a single-zeta plus polarization (SZP) basis set and a mesh cutoff energy of 300 Ry. Brillouin zone sampling was set to $5 \times 5 \times 1$ for graphene and $MoS_2$, and $3 \times 3 \times 3$ for silicon. By setting the *SaveHS* parameter in SIESTA to true, both the Hamiltonian and overlap matrices can be extracted.

**Periodic supercell datasets with defects or dopants.** To evaluate DeepQTH's ability to generalize to disordered systems, we constructed defect-containing datasets using the same simulation parameters as above. These included eight representative point defect configurations in graphene, ten vacancy and substitutional defects in $MoS_2$, and six defect/doping scenarios in silicon (Supplementary Section 8.1). For each structure, we computed the corresponding Hamiltonian and overlap matrices to form a diverse training corpus for learning defect-induced modifications in electronic structure.

**Open device datasets with defects or dopants.** To fine-tune pretrained DeepQTH and train DeepQTV for bias-dependent quantum transport predictions, we generated open-boundary device datasets for graphene, $MoS_2$, and silicon nanostructures. Specifically, we constructed 8 GNR devices, 10 $MoS_2$ nanoribbon devices, and 2 silicon block devices with embedded defects or dopants (Supplementary Section 8.2). MD simulations were conducted at 300 K, keeping the atomic coordinates of the left and right electrodes fixed, while optimizing the scattering region. All simulations employed the same SZP basis and GGA-PBE functional, with a 300 Ry mesh cutoff. Each device class contained 200 stable configurations, partitioned into training, validation, and test sets using a 6:2:2 split. Using TranSIESTA, we performed non-equilibrium calculations with 1 k-point along the transport direction and vacuum directions, 5 k-points in periodic directions. Total potential (set *SaveTotalPotential* parameter to true) as well as Hamiltonian and overlap matrices were computed under multiple bias conditions. The corresponding TPDs, defined as the difference between biased and unbiased total potential, were extracted to train DeepQTV.

Together, these curated datasets form a diverse and rigorous benchmark for evaluating DeepQT's capacity to generalize across different material classes, structural imperfections, and transport regimes, enabling efficient prediction of quantum transport in realistic device scenarios.

**Details of neural network training**

The DeepQT model consists of DeepQTH and DeepQTV, with a separate DeepQT model trained for each material system. The DeepQTH model employs a graph-based Transformer architecture, comprising three GTLs without angular information for the input and hidden layers, followed by a final GTL incorporating angular features as the output layer. The element embedding and node feature dimensions were both set to 64. Edge features were initialized using Gaussian embeddings with a feature dimension of 128 per layer, and a Gaussian broadening parameter σ=0.05. The edge feature dimension was also set to 128. Truncation radii were set to 8.0 Å for graphene, 10.0 Å for monolayer $MoS_2$, and 9.0 Å for silicon to define the local atomic environment.

Each node's structural context was further enriched with Laplacian centrality, degree centrality, and Voronoi-based geometric features, all mapped to 64-dimensional embeddings. Angular information was encoded using spherical harmonics, with the angular quantum number l=5, resulting in a 25-dimensional feature vector capturing angular symmetry. The number of attention heads per GTL layer was set to 4. The output edge feature dimension corresponds to the orbital product space of atomic pairs (i, k), capturing orbital-level interactions. Model training was performed using the Adam optimizer[49] with an initial learning rate of $10^{-3}$, following a stepwise decay strategy that halved the learning rate every 1000 epochs over a total of 3000 epochs.

For the DeepQTV model, the first layer uses GTL with angle information, and then two layers of GTL without angle information are used as the hidden layer and the output layer. Atomic number and potential difference embeddings were each mapped to 64-dimensional embeddings. All other parameters, including edge dimensions, attention mechanisms, and truncation radii, were kept consistent with DeepQTH. The output layer incorporates a 64-dimensional virtual node representation, which is linearly projected to yield scalar TPD values at each mesh point. The optimizer, learning rate schedule, and number of training epochs mirror those used in DeepQTH.

Both DeepQTH and DeepQTV models were implemented using the PyTorch Geometric library[50] and trained on NVIDIA GeForce RTX 4090 GPUs with a batch size of 3. The predicted self-consistent Hamiltonians under bias conditions were subsequently used to compute electronic structure and quantum transport properties. The final transport characteristics were obtained using TBtrans, enabling multi-property prediction and visualization of quantum transport in nano-electronic devices.

**DATA AVAILABILITY**

The data that support the findings of this study are available from the corresponding author upon reasonable request.

**CODE AVAILABILITY**

The code that support the findings of this study are available from the corresponding author upon reasonable request.

**Acknowledgments**

F.L. acknowledges support from National Natural Science Foundation of China under Grant No. T2293700, Grant No. T2293703, Grant No. T2293702, and the 111

Project (B18001). We thank high-performance computing platform of Peking University and National Supercomputer Center in Tianjin for computation facilities.

## Author contributions

F.L. conceived and supervised the project. Z.L.T developed the codes, and performed all calculations Z.L.T and F.L. wrote the manuscript. All the authors contribute to editing the manuscript.

## Competing interests

The authors declare no competing financial interests.

## Additional information

Supplementary information

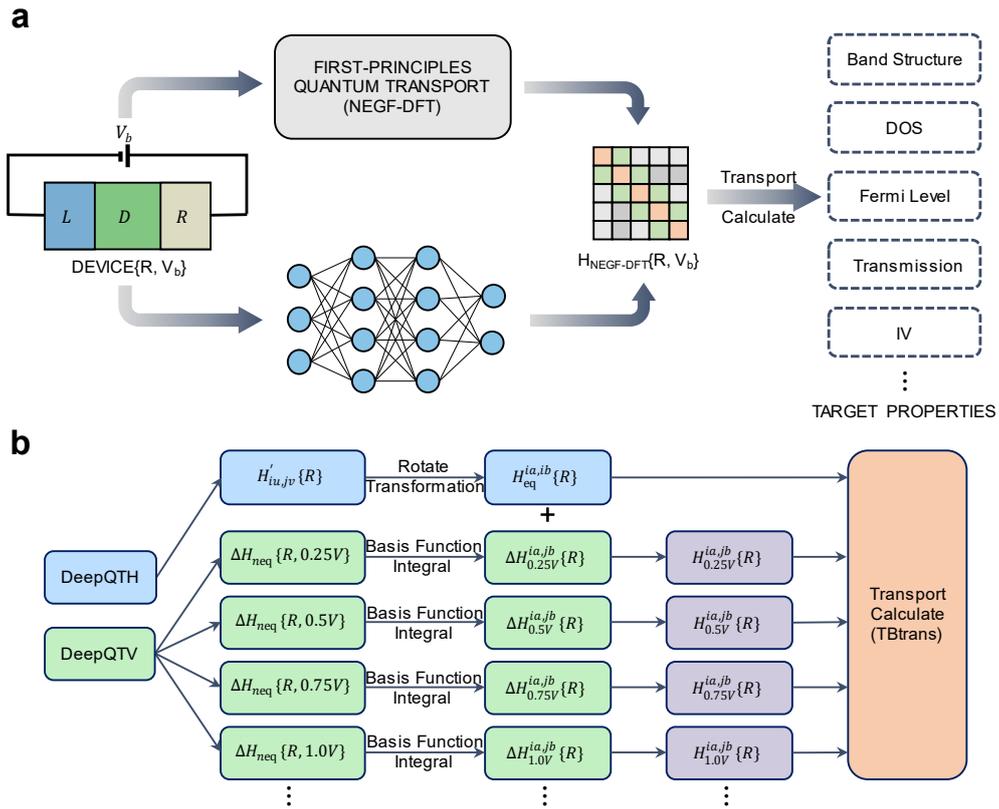

**Figure 1 | Deep learning framework for *ab-initial* quantum transport prediction.**

**a**, Schematic overview of DeepQT, a deep learning framework that bypasses the computationally intensive self-consistent NEGF-DFT procedure. **b**, The DeepQT architecture consists of two integrated sub-models: DeepQTH, which predicts the equilibrium Hamiltonian, and DeepQTV, which estimates the Hamiltonian correction under applied bias. The combined Hamiltonian is then used to compute quantum transport properties via post-processing with TBtrans.

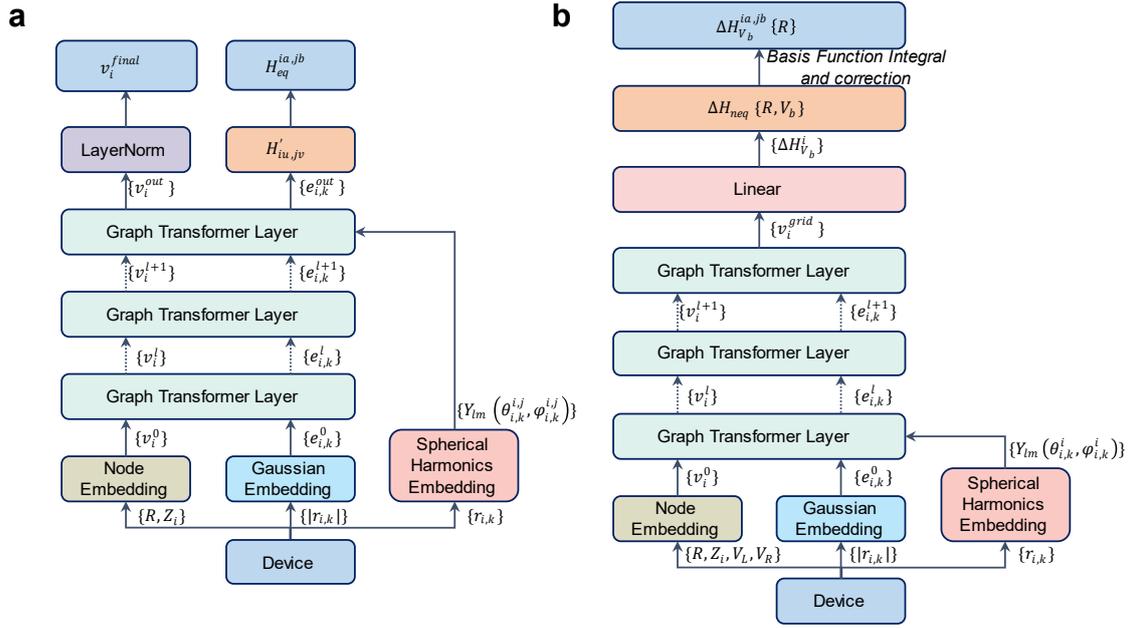

**Figure. 2 Network architectures of the DeepQTH and DeepQTV models.**

**a**, Schematic of the DeepQTH architecture. Atomic structure files are preprocessed into node embeddings, Gaussian edge embeddings, and spherical harmonic azimuthal feature embeddings. Azimuthal information is incorporated only in the output layer. The network extracts atomic-level representations and predicts the equilibrium Hamiltonian. **b**, Schematic of the DeepQTV architecture. The input includes node embeddings for both mesh points and atoms, Gaussian edge embeddings, and azimuthal features, which are integrated into every layer of the Graph Transformer Layer (GTL). The model predicts the total potential difference (TPD), which is subsequently integrated with basis functions to yield the Hamiltonian correction under bias.

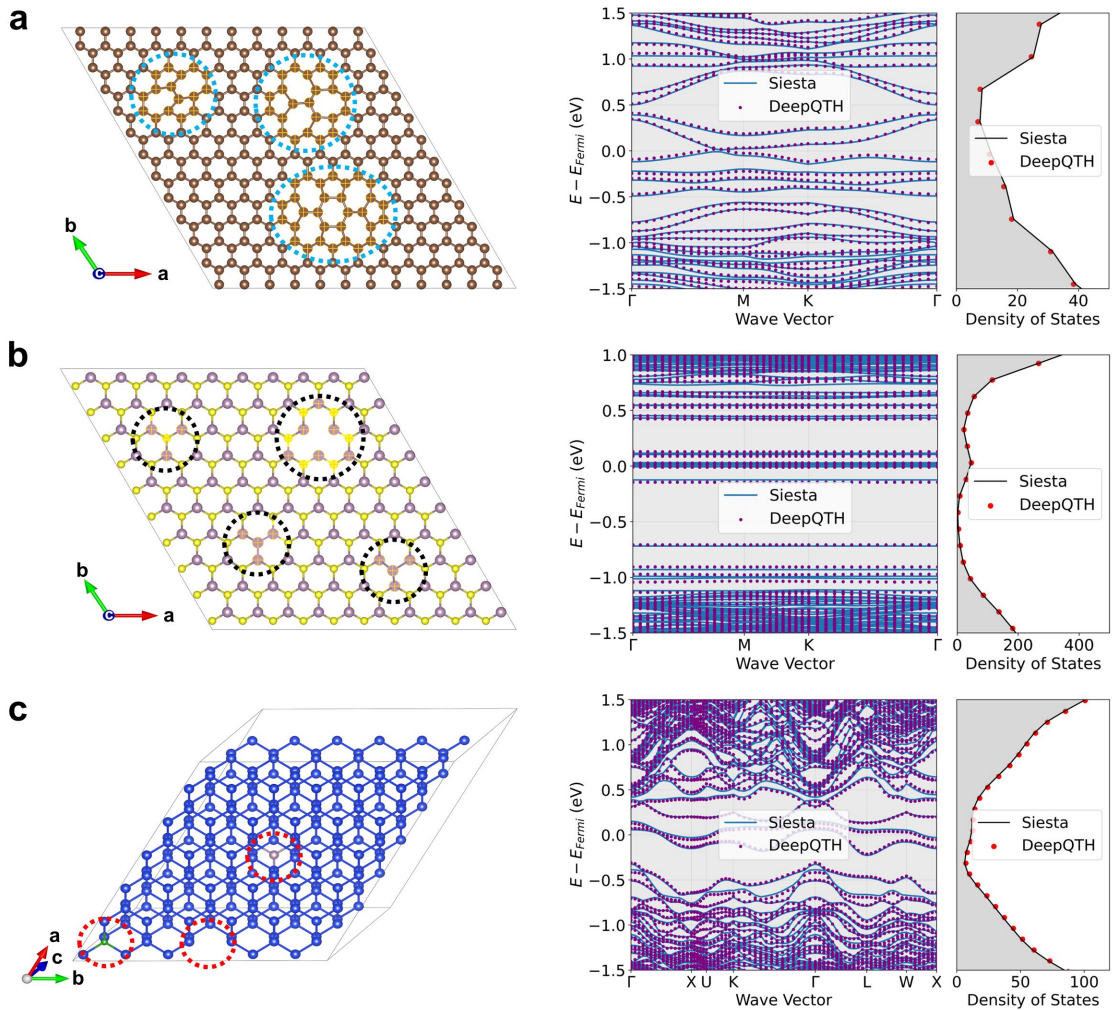

**Figure 3 | DeepQTH prediction of equilibrium electronic structures in large-scale defected and doped systems.**

**a**, Large-scale graphene supercell containing three representative defect types and the corresponding DeepQTH-predicted electronic structure. **b**, $MoS_2$ supercell with four distinct defect configurations and their predicted electronic properties. **c**, Silicon supercell incorporating substitutional dopants and Schottky defects, along with predicted results. These examples highlight the ability of DeepQTH to accurately capture equilibrium electronic structures in complex, defect-rich and doped materials across diverse material systems, demonstrating its robustness and scalability for real-world, large-scale device modeling.

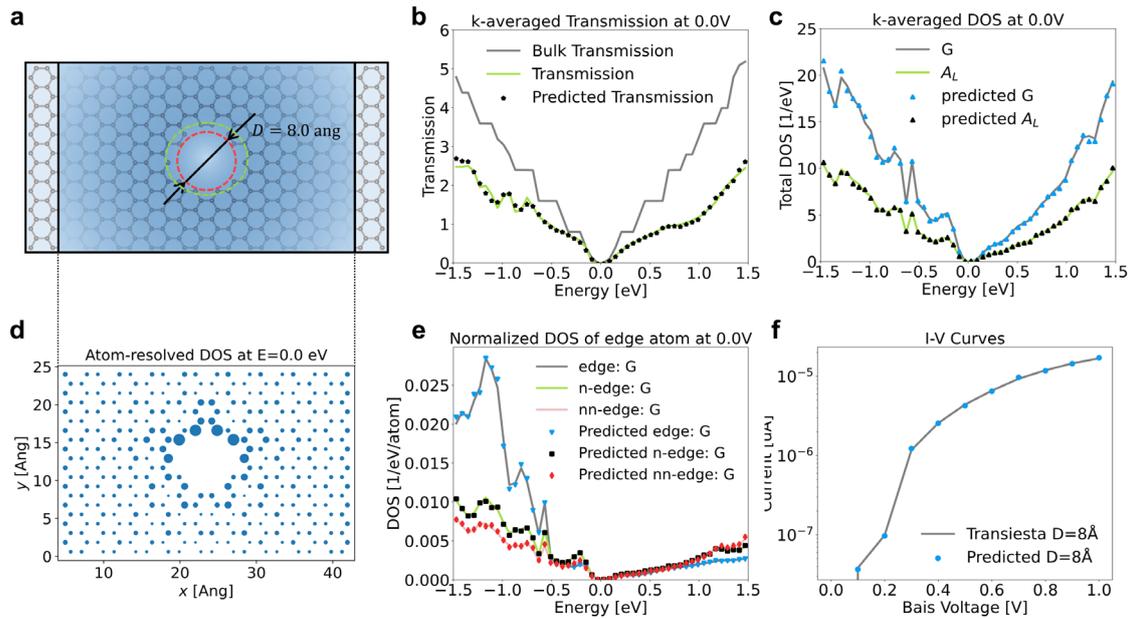

**Figure 4 | DeepQT predictions of quantum transport in large-scale nanoporous graphene nanoribbons (NP-GNRs) with hole defects.**

**a**, Schematic of the NP-GNRs device structure with large-scale defects in the scattering region. **b**, Comparison of DeepQT-predicted and reference k-averaged transmission spectra; "Bulk transmission" denotes the ideal graphene nanoribbon without defects. **c**, Predicted versus calculated spectral density of states for the left electrode and the Green's function density of states in the device. **d**, Atom-resolved density of states (DOS) in the scattering region, highlighting strong localization at the defect edges. **e**, Predicted DOS comparison among edge atoms, nearest neighbors (N-edge), and next-nearest neighbors (NN-edge), showing decay in local states away from the defect. **f**, Predicted and reference current–voltage (*I–V*) characteristics, demonstrating excellent agreement across bias voltages. These results highlight DeepQT's ability to accurately capture defect-induced transport phenomena and localized electronic states in complex large-scale systems.

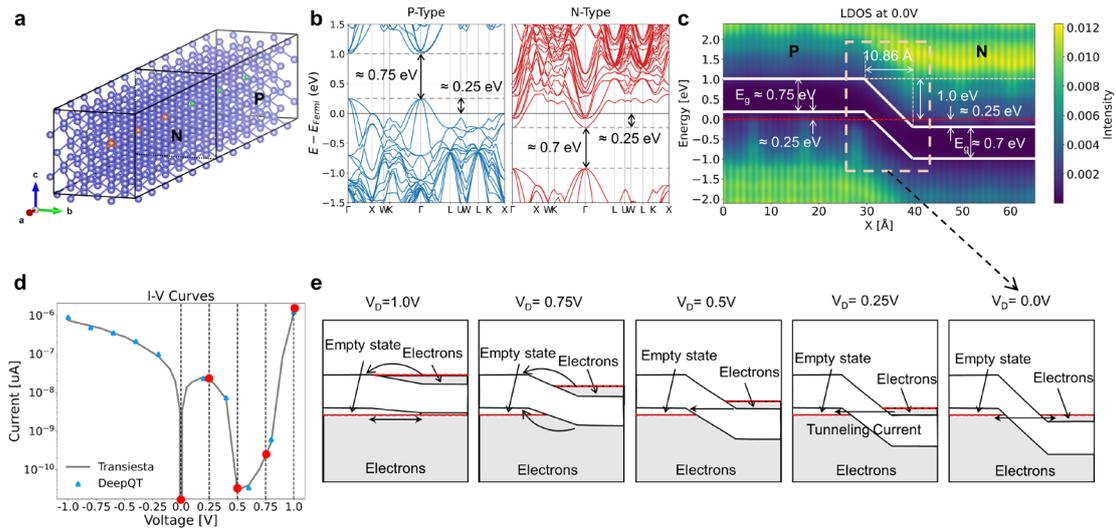

**Figure 5 | Quantum-transport prediction of a heavily doped silicon Esaki diode using DeepQT.**

**a**, Schematic illustration of the heavily doped silicon Esaki diode. **b**, Band structures of the left (p) and right (n) electrodes after Fermi-level alignment computed using SIESTA, showing an upward shift in the p region and a downward shift in the n region. **c**, Zero-bias local density of states (LDOS) revealing a clear band offset and built-in potential barrier; the Fermi level resides in the p-type valence band and n-type conduction band, a hallmark of an Esaki diode. **d**, I-V characteristics computed from DeepQT-predicted Hamiltonians (blue points) agree closely with TranSIESTA calculations (gray curve), capturing nonlinear transport and the negative differential resistance (NDR) between 0.2 and 0.5 V. **e**, Schematic LDOS evolution at selected biases (0.0, 0.25, 0.5, 0.75, and 1.0 V) highlights the tunneling mechanism, revealing the bias-dependent evolution of electronic states and the physical origin of the NDR effect.